\documentclass[twocolumn,showpacs,preprintnumbers,prl,floatfix]{revtex4}
\usepackage{epsfig,amssymb,amsmath,hyperref,bm}
\usepackage{epsfig,amssymb,isolatin1}
\usepackage{graphics}
\usepackage{graphicx}
\usepackage{psfrag}
\newcommand\beq{\begin{equation}}
\newcommand\eeq{\end{equation}}
\newcommand\bea{\begin{eqnarray}}
\newcommand\eea{\end{eqnarray}}
\newcommand{\nonum}{\nonumber}

\begin{document}

\title{On transport in quantum Hall systems with constrictions}

\author{Siddhartha Lal}
\affiliation{The Abdus Salam ICTP, Strada Costiera 11, Trieste 34014, Italy.} 
\email{slal@ictp.it}

\begin{abstract}
Motivated by recent experimental findings, we study transport in a simple 
phenomenological model of a quantum Hall edge system with a gate-voltage 
controlled constriction lowering the local filling factor. The current 
backscattered from the constriction is seen to arise from the matching of 
the properties of the edge-current excitations in the constriction ($\nu_{2}$) 
and bulk ($\nu_{1}$) regions. We develop a hydrodynamic theory for bosonic 
edge modes inspired by this model, finding that a competition between two 
tunneling process, related by a quasiparticle-quasihole symmetry, 
determines the fate of the low-bias transmission conductance.     
In this way, we find satisfactory explanations for many recent puzzling 
experimental results. 
\end{abstract}



\maketitle

The quantum Hall effects are essentially the low temperature physics of a 
disordered 2DEG placed in a strong perpendicular magnetic field 
\cite{yoshioka}. For particular values of the external magnetic field, the 
incompressible Hall fluid in the bulk and gapless current carrying 
edge excitations lead to a vanishing longitudinal resistance and a 
quantised Hall resistance. Electronic correlations are crucial for 
the fractional quantum Hall effect, a gapped ground state with fractionally 
charged quasiparticle excitations \cite{laughlin} which were observed in 
shot-noise measurements \cite{samin}.  Local quasiparticle tunneling 
between the oppositely directed current carrying edge states of a Hall 
bar is known theoretically to be a singular perturbation, with the 
strong-coupling behaviour that of two effectively disconnected 
quantum Hall bubbles \cite{kane}.  Recent experiments studying transport 
through gated constrictions in quantum Hall systems at integer as well as 
fractional filling factors \cite{roddaro,chung} have, however, shown the 
need for a deeper understanding of inter-edge tunneling. 
\par
A signature of departure from the traditional quantum Hall scenario  
can be observed in these experiments from the following. While 
imposing a finite bias at a pair of local split-gates ($V_{G}$) causes 
a backscattered current across the bulk (leading to a finite, edge-bias 
($V$) independent, longitudinal resistance drop), a fractional Hall 
conductance ($g$) is simultaneously measured across the constriction at finite 
$V$ corresponding to a filling factor, $\nu_{2}$, lower than that in the bulk 
($\nu_{1}$). Further, for large $V_{G}$, $g$ is observed to dip sharply 
and vanish with a 
power-law dependence on $V$ as $|V|\rightarrow 0$. A comparison 
with the theory of Fendley etal. \cite{fendley} for inter-edge 
Laughlin quasiparticle tunneling suggests strongly that the constriction 
transmission is governed by $\nu_{2}$.
This is particularly unexpected for an integer quantum Hall 
system \cite{roddaro}. 
\par
A particularly intriguing observation is that of the evolution of 
the transmission conductance $g$ as $V_{G}$ is varied in the limit 
of vanishing $V$. While $g$ shows a 
zero-bias minimum at sufficiently large $V_{G}$, decreasing $V_{G}$ 
leads first to a bias-independent transmission at a particular value of 
the gate-voltage $V_{G}=V_{G}^{*}$ 
and then to an enhanced zero-bias transmission for yet lower values 
of $V_{G}$. This behaviour of the zero-bias $g$ is also 
observed across a wide range of temperatures.
The bias-independence of $g$ at a certain $V_{G}^{*}$ and 
its enhancement at $V_{G}<V_{G}^{*}$ are quite unexpected from the 
conventional theoretical viewpoint \cite{kane}. Qualitatively similar 
results were 
found for the integer cases of $\nu_{1}=1$ \cite{roddaro} as well as for 
the fractional cases of $\nu_{1}=1/3,2/5$ and $3/7$ \cite{roddaro,chung}, 
allowing for the possibility of a common explanation. 
\begin{figure}[htb]
\begin{center}
\scalebox{0.4}{
\psfrag{1}[bl][bl][4][0]{1}
\psfrag{2}[bl][bl][4][0]{2}
\psfrag{3}[bl][bl][4][0]{3}
\psfrag{4}[bl][bl][4][0]{4}
\psfrag{5}[bl][bl][4][0]{G}
\psfrag{6}[bl][bl][4][0]{G}
\psfrag{7}[bl][bl][4][0]{S}
\psfrag{8}[bl][bl][4][0]{D}
\psfrag{9}[bl][bl][4][0]{$\nu_{1}$}
\psfrag{10}[bl][bl][4][0]{$\nu_{1}$}
\psfrag{11}[c][c][4][0]{$\nu_{2}$}
\includegraphics{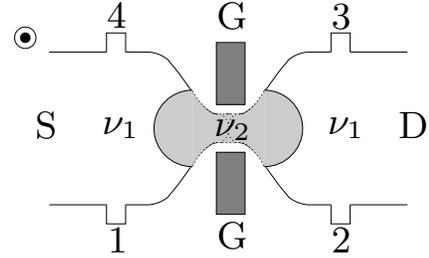}}
\end{center}
\caption{A schematic diagram of a QH bar at a bulk filling fraction $\nu_{1}$ 
and with a gate-voltage (G) 
controlled split-gate constriction which lowers the filling 
fraction in the constriction region to $\nu_{2}$.
S and D show the source and drain ends  
of the Hall bar while 1 to 4 signify the current/voltage terminals.} 
\label{kf1}
\end{figure}
\par
The most probable effect of a split-gate system is to create a 
smooth and long constriction potential, depleting the local electronic 
density (and hence lowering the local filling factor) locally from its 
value in the bulk. This led Roddaro 
and co-workers \cite{roddaro} to conjecture on the likelihood of a 
small region in the neighbourhood of the constriction with a reduced 
filling factor ($\nu_{2}<\nu_{1}$) being the cause of the puzzling results 
mentioned above (see Fig.(\ref{kf1})). In what follows, we 
develop a simple phenomenological model based on this conjecture and aim to 
provide satisfactory explanations for the experimental results. In doing 
so, we devote our attention solely to short-ranged electronic correlations 
which cause the formation of chiral Tomonaga-Luttinger liquid (TLL) edge 
states without the intervention of any stripe states \cite{papastroh} 
arising from longer range interactions. Further, the 
model neglects any line-junction non-chiral TLLs formed across the 
split-gates \cite{papamac} and effects of inter-edge interactions on 
quasiparticle tunneling \cite{agosta}, focusing instead on the 
transmitted and reflected edge states arising from the nature of the 
Hall fluid inside the constriction. 
\par  
We begin by deriving some results for ballistic edge transport through a 
constriction region with reduced filling fraction $\nu_{2}$ in a Hall bar 
geometry (see Fig.(\ref{kf1})) from a few simple considerations. 
We make two reasonable assumptions in framing the model. 
First, that the voltage (Hall) bias between the two edges of the sample 
is not affected 
by the local application of a gate-voltage at a constriction as long 
as the bulk of the system is in an incompressible quantum Hall state 
($\nu_{1}$). 
Second, that the two-terminal conductance measured across the constriction 
is determined by the 
lowered filling-fraction of the quantum Hall ground state in the 
constriction $(\nu_{2})$.
\par
Then, for a current $I$ injected from the source, 
we know that $I = g_{b} V_{42}$ where 
$g_{b}=\nu_{1}e^{2}/h$ is the bulk Hall conductance and $V_{42}$ is 
the source-drain edge-bias. The second assumption gives the current 
transmitted ballistically through the constriction as  
$I_{tr} = g_{c} V_{42}$, where $g_{c}=\nu_{2}e^{2}/h$ is the two-terminal 
conductance measured across the constriction. 
From the first assumption, we then obtain 
the transmitted current $I_{tr}$ as 
\beq
I_{tr} = \frac{g_{c}}{g_{b}} I = \frac{\nu_{2}}{\nu_{1}} I~.
\label{transcurr}
\eeq
From Kirchoff's law for current conservation, we get the current 
reflected at the constriction as
$I_{ref} = I - I_{tr} = (1 - \nu_{2}/\nu_{1}) I$. This, in turn, gives the 
minimum value of the backscattering conductance as 
\beq
g^{back}=I_{ref}/V_{42} = (1 - \nu_{2}/\nu_{1}) G_{b} 
= (\nu_{1} - \nu_{2})\frac{e^{2}}{h}~.
\label{backcond}
\eeq 
and is seen to be quantised at an effective filling factor for the 
reflected current as 
$\nu_{ref} = \nu_{1} - \nu_{2}$ (in units of $e^{2}/h$). 
Further, we find the ``background", edge-bias independent, value of 
the longitudinal resistance drop across the constriction to be 
\beq
R^{BG}=\frac{V_{4} - V_{3}}{I} 
= (1 - \frac{\nu_{2}}{\nu_{1}}) g_{b}^{-1}~.
\label{bgroundres}
\eeq
\par
\begin{figure}[htb]
\begin{center}
\scalebox{0.4}{
\psfrag{1}[bl][bl][3][0]{$1,in$}
\psfrag{2}[bl][bl][3][0]{$1,out$}
\psfrag{3}[bl][bl][3][0]{$2,in$}
\psfrag{4}[bl][bl][3][0]{$2,out$}
\psfrag{5}[bl][bl][3][0]{$u$}
\psfrag{6}[bl][bl][3][0]{$r$}
\psfrag{7}[bl][bl][3][0]{$d$}
\psfrag{8}[bl][bl][3][0]{$l$}
\psfrag{9}[bl][bl][3][0]{$\nu_{2}$}
\psfrag{10}[bl][bl][3][0]{$\nu_{1}$}
\psfrag{11}[bl][bl][3][0]{$\nu_{1}$}
\psfrag{12}[c][c][2.5][0]{$(x\sim 0)$}
\includegraphics{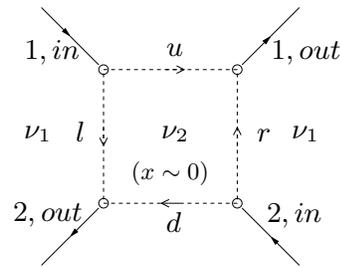}
}
\end{center}
\caption{A schematic diagram of the ``constriction" system given 
by the dashed box around the region $x\sim 0$ and symbolised by the filling 
fraction $\nu_{2}$ lower than that of the bulk, $\nu_{1}$. The 
four chiral fields approaching and leaving this region are shown by 
the arrows marked as $1,in$, $1,out$, $2,in$ and $2,out$. The dashed 
horizontal and vertical lines at the junction represent the edge states 
which are transmitted $(u,d)$ and reflected $(l,r)$
at the constriction respectively.} 
\label{beam1}
\end{figure}

The current backscattered from the constriction is presumably carried 
in a gapless region lying in-between the bulk and constriction regions. By 
relying on the same assumptions, we 
now present a hydrodynamic model of gapless, current carrying, chiral 
edge density-wave
excitations, similar in spirit to the classic work of Wen \cite{wen}, 
describing ballistic transport through the transmitting and reflecting 
edges at the constriction (shown schematically in Fig.(\ref{beam1})). 
As the model is an effective field 
theory quadratic in the bosonic fields describing the edge excitations, 
inter-edge quasiparticle tunneling will be seen to arise from the 
exponentiation of these fields. We will focus here on presenting the 
results of the model and their immediate relevance to the experiments, 
keeping the mathematical details for elsewhere \cite{inprep}. 
\par
The extent of the constriction region $2a$ is assumed to lie in the range 
$l_{B}<<2a<<L$, 
where $L$ is the total system size and $l_{B}$ is the magnetic length, with 
the external arms $(1in,\ldots,2out)$ meeting the internal ones $(u,\ldots,l)$ 
at the 4 corners of the constriction. In keeping with the assumptions, the 
filling factor governing the properties of the four outer arms is $\nu_{1}$ 
and those of the upper and lower arms of the circuit at the constriction are 
governed by $\nu_{2}$. The effective filling-factor governing the properties 
of the right and left arms of the circuit is treated as a parameter 
$\nu_{ref}$ whose value will be determined from the analysis.
Ballistic transport of current carrying edge excitations in the various 
arms of the circuit shown in Fig.(\ref{beam1}) is given by 
a Hamiltonian $H$ describing the energy cost for edge-density distortions 
\cite{wen} $H = H^{ext} + H^{int}$ where
\beq
H^{ext} = \frac{\pi v}{\nu_{1}}[\int_{-L}^{-a}\hspace*{-0.5cm}
dx~(\rho_{1in}^{2}
+\rho_{2out}^{2}) + \int_{a}^{L}\hspace*{-0.3cm}
dx~(\rho_{2in}^{2}+\rho_{1out}^{2})]
\eeq
and $H^{int}$ has the same form as $H^{ext}$ but with the densities 
$(\rho_{u},\ldots,\rho_{l})$ and filling factors $\nu_{2}$ and $\nu_{ref}$ 
placed appropriately. We have taken the velocity $v$ of the edge-excitations 
to be the same for all arms, focusing instead on the effects of a 
changing 
filling factor. The densities $\rho$ are, as usual, represented in terms 
of bosonic fields $\phi$ describing the edge displacement, e.g, 
$\rho_{1in}= 1/2\pi \partial_{x}\phi^{1in}$, 
$\rho_{2out} =-1/2\pi \partial_{x}\phi^{2out}$ \cite{wen}. The commutation 
relations satisfied by these fields are familiar, e.g.,
\beq
[\phi^{1in}(x),\partial_{x}\phi^{1in}(x')] = i\pi\nu_{1}\delta(x-x')
\eeq
and so on for the other fields. The equations of motion found from $H$ 
describe the ballistic motion of chiral edge-density 
waves, e.g, $(\partial_{t} + v\partial_{x})\rho^{1in}(x,t)=0$, 
$\rho^{1in}(x,t)\equiv\rho^{1in}(x-vt)$ etc. The $H$ given above, 
however, needs to be supplemented with matching conditions at the four corners 
of the constriction in order to give a complete description. From the form 
of $H$, it is clear that we need two matching conditions at 
each corner; a reasonable choice is one defined on the fields and 
one on their derivatives. For the sake of brevity, we present here only 
those at the top-left corner 
\bea
\phi^{1in}(x=-a) &=& \phi^{u}(x=-a) + \phi^{l}(y=-a)\nonum\\
\partial_{x}\phi^{1in}(x=-a) &=& \partial_{x}\phi^{u}(x=-a) 
+ \partial_{y}\phi^{l}(y=-a) 
\label{match}
\eea 
where $x$ and $y$ are the spatial coordinates describing the $(1in,u)$ 
and $l$ arms respectively. The equation of continuity leads to the familiar 
form for the current operator, e.g, $j^{1in}=1/2\pi \partial_{t}\phi^{1in}$ 
etc. Thus, by taking the time derivative throughout the first of the 
matching conditions (\ref{match}), we see that current conservation is 
guaranteed at every corner. Further, from the two matching conditions, we 
compute the commutation relation
\beq
[\phi^{l},\partial_{y}\phi^{l}]_{y\rightarrow-a}
=([\phi^{1in},\partial_{x}\phi^{1in}]-
[\phi^{u},\partial_{x}\phi^{u}])_{x\rightarrow-a}
\eeq
which immediately leads to the effective filling factor for the reflected 
current arms, derived earlier from simpler considerations, as 
$\nu_{ref}=\nu_{1}-\nu_{2}$. Another check involves computing the 
chiral conductances $g_{1in,1out}$ and $g_{1in,2out}$ (where the first 
and second indices give the incoming and outgoing current carrying 
arms respectively) in the presence of a small, finite source-drain bias. 
Employing the standard Kubo formulation relating conductance to a 
current-current correlator \cite{giamarchi},  
we reproduce the results 
$g_{1in,1out}= \nu_{2}$ and $g_{1in,2out}=\nu_{1}-\nu_{2}$ (in units of 
$e^{2}/h$).   
\par
We now account for 
the role of quasiparticle tunneling in determining low-energy 
transport. First, it is clear that local quasiparticle (qp)
tunneling processes between the $(u,d)$ arms will be dictated by the 
constriction filling factor $\nu_{2}$. We can write such a tunneling term, 
located deep inside the constriction at $x=0$, as 
$\lambda_{1}\cos(\phi^{u}(x=0)-\phi^{d}(x=0))$. From the 
work of Kane and Fisher \cite{kane}, we know that the RG equation for 
the qp tunnel coupling $\lambda_{1}$ is given by
$d\lambda_{1}/dl=(1-\nu_{2})\lambda_{1}$~.
With $\nu_{2}<1$, $\lambda_{1}$ will grow under the RG flow to 
strong coupling.
\par
\begin{figure}[htb]
\begin{center}
\scalebox{0.3}{
\psfrag{1}[bl][bl][5][0]{$\frac{\nu_{2}}{\nu_{1}}$}
\psfrag{2}[bl][bl][4][0]{$0$}
\psfrag{3}[bl][bl][4][0]{$0$}
\psfrag{4}[bl][bl][4][0]{$1$}
\psfrag{5}[bl][bl][4][0]{$1$}
\psfrag{6}[bl][bl][4][0]{(a)}
\psfrag{7}[bl][bl][4.9][0]{$1-\frac{\nu_{2}}{\nu_{1}}$}
\psfrag{8}[bl][bl][4][0]{$1$}
\psfrag{9}[bl][bl][4][0]{$1$}
\psfrag{10}[bl][bl][4][0]{$0$}
\psfrag{11}[bl][bl][4][0]{$0$}
\psfrag{12}[c][c][4][0]{(b)}
\psfrag{13}[c][c][4][0]{(c)}
\psfrag{14}[c][c][4][0]{(d)}
\includegraphics{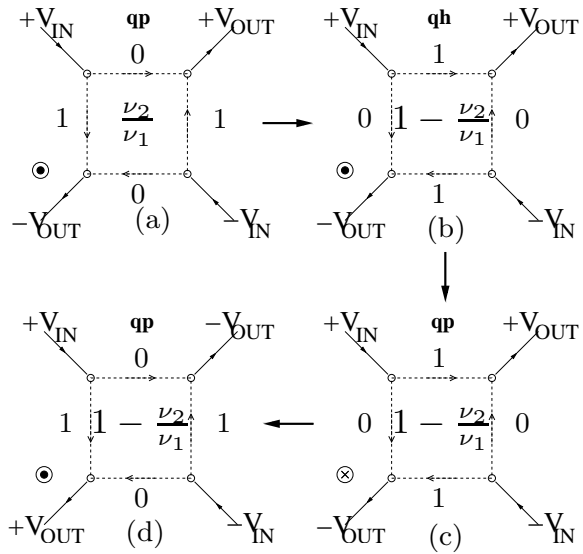}
}
\end{center}
\caption{The quasiparticle-quasihole (qp-qh) symmetry of the 
$(\nu_{1},\nu_{2})$ constriction geometry in terms of the relative filling 
factor. A similar diagram for the $\nu_{1}=1$ system was presented in 
Ref.(\cite{roddaro}). The source-drain bias $2V_{in}$ is applied to the 
two incoming arms while $\pm V_{out}$ are the equilibration potentials 
of the two outgoing arms. We first map the original circuit (a) of qps 
onto that of qhs (b) via a qp-qh conjugation, followed by mapping onto 
a circuit of qps with the external magnetic field reversed (c) and finally 
a mapping onto a circuit of qps with the original geometry (d) by rotating 
around the axis of the two outgoing arms by $180^{\circ}$. In this way, an 
equivalence is established between the circuits in (a) and (d).}
\label{qpqh}
\end{figure}

 
There is, however, also a tunneling process between the $(l,r)$ arms to 
consider. It is revealed by the quasiparticle-quasihole (qp-qh) symmetry of 
the completely filled effective lowest Landau level of the qps (i.e., the 
ground state of the Hall fluid) in the bulk which is protected by a gap 
larger than all other energy scales in the problem \cite{girvin}. A similar 
argument for the electron-hole symmetry of the $(\nu_{1}=1,\nu_{2})$ 
constriction geometry was presented in 
Ref.(\cite{roddaro}); in Fig.(\ref{qpqh}), we extend it to a general 
$(\nu_{1},\nu_{2})$ 
system by employing the notion of a relative filling factor (obtained 
by dividing throughout by the bulk $\nu_{1}$).
\par
A visual representation of the steps in the argument is presented in 
Fig.(\ref{qpqh}). For the system as shown in Fig.(\ref{beam1}), 
we have a relative filling factor of 
unity in the bulk and that of $\nu_{2}/\nu_{1}$ of the Hall fluid inside the 
constriction (Fig.(\ref{qpqh}a)). Now, a partially filled effective 
lowest Landau level of qps with relative filling $\nu_{2}/\nu_{1}$ can 
equivalently be studied in terms of a partially filled effective Landau 
level of qhs with relative filling $1-\nu_{2}/\nu_{1}$ over a completely 
filled effective Landau level of qps. Thus, we can carry out a qp-qh 
conjugation transformation (Fig.(\ref{qpqh}b)) to go to a description 
in terms of qhs. Noting that qhs are time-reversed qps, we can map 
the qh system onto that of a qp system with the direction of the external 
magnetic field reversed (Fig.(\ref{qpqh}c)). A final rotation 
of $180^{\circ}$ about the axis of the two outgoing arms (Fig.(\ref{qpqh}d)) 
leaves us with a circuit of qps with the same geometry. Two things, 
however, have changed in undertaking 
this series of transformations. First, the relative filling of the 
constriction has changed from $\nu_{2}/\nu_{1}$ to $1-\nu_{2}/\nu_{1}$. 
Second, the transmitted and reflected outgoing arms (defined with respect 
to the source-drain bias) of the original circuit have been interchanged in 
reaching the final one; the transmitted and reflected conductances of the two 
circuits are now, in fact, linked by duality relations \cite{roddaro,inprep}. 
Equally importantly, in analogy with the qp tunneling process between 
the two transmitted current arms $(\phi^{u},\phi^{d})$, this 
qp-qh transformation reveals the existence of a 
qh-tunneling process between the two reflected current arms 
$(\phi^{l},\phi^{r})$ of the constriction: 
$\lambda_{2}\cos(\phi^{l}(y=0)-\phi^{r}(y=0))$, 
with the RG equation for the qh-tunnel coupling $\lambda_{2}$ governed 
by the relative filling $1-\nu_{2}/\nu_{1}$ 
\beq
\frac{d\lambda_{2}}{dl}=(1-(1-\frac{\nu_{2}}{\nu_{1}}))\lambda_{2}
=\frac{\nu_{2}}{\nu_{1}}\lambda_{2}~.
\eeq
Again, with $\nu_{2}<\nu_{1}$ and $(\nu_{1},\nu_{2})>0$, $\lambda_{2}$ will 
also grow under the RG flow to strong coupling.
\par
\begin{figure}[htb]
\begin{center}
\scalebox{0.3}{
\psfrag{1}[bl][bl][4][0]{$\ln\lambda_{1}$}
\psfrag{2}[bl][bl][4][0]{$\ln\lambda_{2}$}
\psfrag{3}[bl][bl][4][0]{$\nu_{2}<\nu_{2}^{*}$}
\psfrag{4}[bl][bl][4][0]{$\nu_{2}>\nu_{2}^{*}$}
\psfrag{5}[bl][bl][4][0]{$\nu_{2}=\nu_{2}^{*}$}
\includegraphics{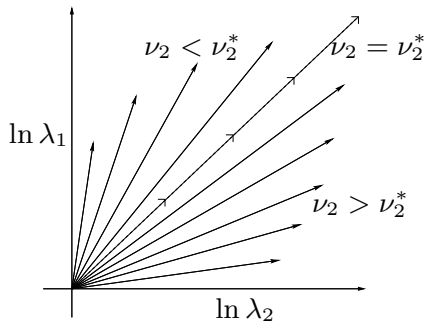}
}
\end{center}
\caption{The RG phase diagram for the model as a plot of the 
function $\ln\lambda_{1}/\ln\lambda_{2} = \nu_{1}(1/\nu_{2} - 1)$. 
All RG flows lead away from the weak-coupling unstable fixed point at 
the origin. The dashed line with unit slope represents the critical case 
of $\nu_{2}=\nu_{2}^{*}$, where qp-qh symmetry fixes the constriction 
transmission at its weak coupling value. The region above (below) the 
critical line contains all RG flows with slopes greater (lesser) than 
unity and $\nu_{2}<\nu_{2}^{*}$ ($\nu_{2}>\nu_{2}^{*}$), for which 
$\lambda_{1}$ ($\lambda_{2}$) grows to strong coupling faster than 
$\lambda_{2}$ ($\lambda_{1}$), causing a dip (peak) in the transmission.} 
\label{rgph}
\end{figure}

However, the couplings 
$\lambda_{1}$ and $\lambda_{2}$ affect the transmission through the 
constriction in opposite ways: while $\lambda_{1}$ reduces 
the constriction transmission, $\lambda_{2}$ increases it. A 
comparison of the two RG equations reveals that for a critical value of 
$\nu_{2}^{*}=\nu_{1}/(1+\nu_{1})$, both couplings grow equally quickly 
and the qp-qh symmetry of the system fixes 
the constriction transmission $t$ (and hence also the reflection) 
at its weak-coupling value of 
$t(\nu_{2}^{*})=g_{1in,1out}/\nu_{1}=1/(1+\nu_{1})$. The critical 
$(\nu_{2}^{*},t(\nu_{2}^{*}))$ values obtained for $\nu_{1}=1,1/3$ are 
$(1/2,1/2)$ and $(1/4,3/4)$ respectively; these 
match exactly the critical filling factor and associated bias-independent 
transmission values obtained in Ref.(\cite{roddaro}). Further, we see 
that for $\nu_{2}<\nu_{2}^{*}$ ($\nu_{2}>\nu_{2}^{*}$), the coupling 
$\lambda_{1}$ ($\lambda_{2}$) will grow to strong coupling faster than 
the coupling $\lambda_{2}$ ($\lambda_{1}$), thereby causing a
dip (peak) in the constriction transmission at low energies (bias/temperature). 
This is 
in conformity with the puzzling zero-bias evolution of the constriction 
transmission with the gate voltage. 
This is also reflected in 
the chiral conductances $g_{1in,1out} \rightarrow 0$ ($\nu_{1}$) and 
$g_{1in,2out} \rightarrow \nu_{1}$ ($0$) in the strong coupling limit 
of $\nu_{2}<\nu_{2}^{*}$ ($\nu_{2}>\nu_{2}^{*}$). 
These results are summarised in the RG phase diagram for our model 
(Fig.(\ref{rgph})), as a plot of the 
function $\ln\lambda_{1}/\ln\lambda_{2} = \nu_{1}(1/\nu_{2} - 1)$. 
\par
In conclusion, we have demonstrated that the puzzling results obtained 
in the experiments \cite{roddaro,chung} can be understood from 
a simple model for the constriction with a filling factor 
lower than that of the bulk. Ballistic 
transport in the presence of a finite backscattered current (and 
associated longitudinal resistance) is understood as a consequence of 
current conservation and matching of the properties of 
the edge excitations in the bulk and constriction regions. By invoking a 
qp-qh symmetry of the system, we explain the observed evolution of the 
low-bias constriction transmission as arising from the competition between 
a qp and a qh tunneling process in determining the conductances at strong 
coupling, as well as make predictions for the critical constriction filling 
factor $\nu_{2}^{*}$ and associated 
constriction transmission $t(\nu_{2}^{*})$ for any quantum Hall system 
with such a constriction.       
\begin{acknowledgments}
I am grateful to D. Sen and S. Rao for many stimulating discussions and 
constant encouragement. Special thanks are due to A.  Altland, 
B. Rosenow, Y.Gefen and R. Mazzarello for many invaluable discussions. I 
am indebted to CCMT, IISc (Bangalore) and HRI 
(Allahabad) for their hospitality while this work was carried out. 
\end{acknowledgments}


\begin{thebibliography}{99}

\bibitem{yoshioka} D. Yoshioka, {\it The Quantum Hall Effect} (Springer 
Series in Solid-State Sciences no. 133, Berllin, 2002) and references therein.

\bibitem{laughlin} R. B. Laughlin, Phys. Rev. Lett. {\bf 50}, 1395 (1983).

\bibitem{samin} L. Saminadayar {\it et al.}, Phys. Rev. Lett. {\bf 79}, 
2526 (1997); R. de Picciotto {\it et al.}, Nature {\bf 389}, 162 (1997).

\bibitem{kane} C. L. Kane and M. P. A. Fisher, Phys. Rev. Lett. {\bf 68}, 
1220 (1992); Phys. Rev. B {\bf 46}, 15233 (1992); X. G. Wen, Int. J. of 
Mod. Phys. B {\bf 6}, 1711 (1992) .

\bibitem{roddaro} S. Roddaro, V. Pellegrini, F. Beltram, L. N. Pfeiffer 
and K. W. West, Phys. Rev. Lett. {\bf 95}, 156804 (2005); S. Roddaro, 
V. Pellegrini, F. Beltram, G. Biasiol and L. Sorba, Phys. Rev. Lett. 
{\bf 93}, 046801 (2004).

\bibitem{chung} Y. C. Chung, M. Heiblum and V. Umansky, Phys. Rev. Lett. 
{\bf 91}, 216804 (2003).

\bibitem{fendley} P. Fendley, A. W. W. Ludwig and H. Saleur, Phys. Rev. 
Lett. {\bf 74}, 3005 (1995); Phys. Rev. B {\bf 52}, 8934 (1995).

\bibitem{papastroh} E. Papa and T. Stroh, Phys. Rev. Lett {\bf 97}, 046801 
(2006); {\it ibid.}, cond-mat/0607458.

\bibitem{papamac} E. Papa and A. H. MacDonald, Phys. Rev. Lett. {\bf 93}, 
126801 (2004); {\it ibid.}, Phys. Rev. B {\bf 72}, 045324 (2005).

\bibitem{agosta} R. D'Agosta, R. Raimondi and G. Vignale, Phys. Rev B 
{\bf 68}, 035314 (2003); L. P. Pryadko, E. Shimshoni and A. Auerbach, 
Phys. Rev. B {\bf 61}, 10929 (2000). 

\bibitem{wen} See X. G. Wen, Advances in Physics {\bf 44}, 405 (1995) for an  
excellent review.

\bibitem{inprep} S. Lal, in preparation.

\bibitem{giamarchi} Thierry Giamarchi, {\it Quantum Physics in One Dimension} 
(Oxford University Press, Oxford, 2004); C. L. Kane and M. P. A. Fisher in 
{\it Perspectives in Quantum Hall Effects}, S. Das Sarma and A. Pinczuk eds., 
John Wiley Publ., New York (1997).

\bibitem{girvin} S. M. Girvin, Phys. Rev. B {\bf 29}, 6012 (1984). 

\end{thebibliography}
\end{document}